\documentclass[%
twocolumn,
showpacs,preprintnumbers,
 aps,
prstab,
]{revtex4-1}
\usepackage{graphics} % for pdf, bitmapped graphics files
\usepackage[pdftex]{graphicx}
\usepackage{epstopdf}
\usepackage[cmex10]{amsmath}
\usepackage{hepparticles}
\usepackage{amssymb}  % assumes amsmath package installed
\usepackage{footnote}
\usepackage{xcolor}
\usepackage[normalem]{ulem}

\begin{document}

% Use the \preprint command to place your local institutional report
% number in the upper righthand corner of the title page in preprint mode.
% Multiple \preprint commands are allowed.
% Use the 'preprintnumbers' class option to override journal defaults
% to display numbers if necessary
%\preprint{}

%Title of paper
%\title{Feedback for Optimization of Betatron Oscillation Minimization of a Time-Varying Lattice}

\title{Nuclear recoil spectroscopy of levitated particles}

% repeat the \author .. \affiliation  etc. as needed
% \email, \thanks, \homepage, \altaffiliation all apply to the current
% author. Explanatory text should go in the []'s, actual e-mail
% address or url should go in the {}'s for \email and \homepage.
% Please use the appropriate macro foreach each type of information

% \affiliation command applies to all authors since the last
% \affiliation command. The \affiliation command should follow the
% other information
% \affiliation can be followed by \email, \homepage, \thanks as well.

\author{Alexander~Malyzhenkov\thanks{qqq}}
\altaffiliation[Also at the ]{Department of Physics, Northern Illinois University, Dekalb, IL, USA}
\affiliation{Chemistry Division, Los Alamos National Laboratory, Los Alamos, NM, USA}

\author{Vyacheslav~Lebedev}
\affiliation{Chemistry Division, Los Alamos National Laboratory, Los Alamos, NM, USA}

\author{Alonso~Castro}
\email[Corresponding author: ]{acx@lanl.gov}
\affiliation{Chemistry Division, Los Alamos National Laboratory, Los Alamos, NM, USA}

%\thanks{This research was supported by Los Alamos National Laboratory.}

%Collaboration name if desired (requires use of superscriptaddress
%option in \documentclass). \noaffiliation is required (may also be
%used with the \author command).
%\collaboration can be followed by \email, \homepage, \thanks as well.
%\collaboration{}
%\noaffiliation

\date{\today}

\begin{abstract}
We propose a new method for the detection and characterization of nuclear decay processes. Specifically, we describe how nuclear decay recoil can be observed within small particles levitated in an optical trap with high positional resolution. Precise measurements of the magnitude of each recoil as well as their rate of occurrence can provide accurate information about the isotopic composition of a radioactive sample. We expect that this new technique for nuclear material characterization will be especially useful in the area of nuclear forensic analysis.
\end{abstract}

% insert suggested PACS numbers in braces on next line
% insert suggested keywords - APS authors don't need to do this
%\keywords{}

%\maketitle must follow title, authors, abstract, \pacs, and \keywords
\maketitle

\section{Introduction}
Nuclear decay is a key phenomenon of nature, spanning various fields of scientific study, such as nuclear physics, radiochemistry, radiobiology and medicine. The measurement of radioactivity in a sample and the determination of the type of emitted nuclear particles is essential for these fields because they allow the characterization of the isotopic composition of the sample. In principle, knowledge of the decay energy and the emission rate unambiguously identifies the specific decaying isotope. There are many techniques currently in use for the detection and characterization of the products of nuclear decay: alpha spectrometry, gamma spectrometry, beta detection, scintillation counting, cloud chamber detection, etc. A more general analytical technique, mass spectrometry, is also widely used for the isotopic analysis of nuclear materials, consisting of ionizing chemical species and sorting the ions based on their mass-to-charge ratio~\cite{Lee}.

In this paper, we consider a conceptually different and new approach to nuclear sample analysis by examining the recoil of a daughter atom within a solid particle of the material caused by each individual decay event. Specifically, we consider small particle samples which are difficult to analyze with the above-mentioned methods. Small particle analysis is important for environmental monitoring, and especially in nuclear forensics, in which the isotope ratios of nuclear materials present in individual particles are measured in swipe samples taken from the inside and outside of nuclear facilities~\cite{Lee}. We show that, for micrometer and sub-micrometer particles, the kinetic energy of the emitted nuclear particle can be determined very accurately by measuring the recoil of the sample particle levitated in an optical trap. The recoil momentum of the daughter atom is fully absorbed by the sample particle that contains it, resulting in a well defined oscillation in the harmonic potential of the optical trap. We demonstrate that this motion due to nuclear recoil can be measured by currently available state-of-the-art techniques in optical trapping~\cite{T_Li_Brownian, T_Li, Novotny, Novotny1, Geraci, GrattaTrap}. Moreover, we calculate that an experimental setup similar to those previously described \cite{ T_Li, Novotny} would be able to measure such recoils with high resolution, thus enabling a new method for the characterization of nuclear samples. 

In section~\ref{TM} we present the theoretical model that describes the micro-particle recoil in an optical trap due to nuclear decay. In section~\ref{limits} we explain the fundamental and practical limits for the detection of nuclear recoil of an optically trapped particle. In section~\ref{spectroscopy} we examine the possibility of using this technique as a nuclear recoil spectrometer. We discuss potential applications of the proposed method in section~\ref{discuss}.

\section{Theoretical model}\label{TM}

%\subsection{Concept: fundamental law of momentum conservation is the key!}
\begin{figure}[h]
	\centerline{\includegraphics[width=0.49\textwidth]{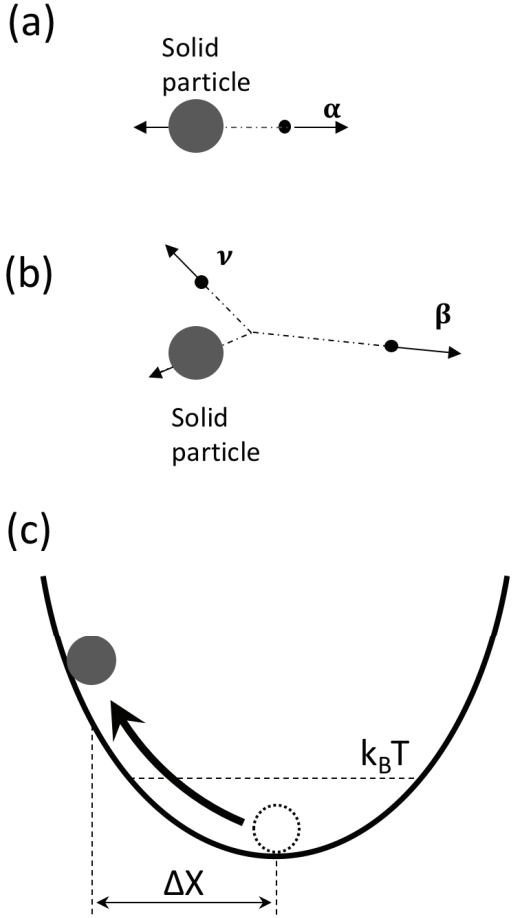}}
	\caption{(a) Schematic diagram of $\alpha-$decay in a free solid particle: the solid particle containing the daughter atom (not pictured) recoils in the opposite direction to that of the alpha particle. (b) Schematic diagram of $\beta-$decay in a free solid particle: a neutrino ($\nu$) and a beta particle initiate recoil of the solid particle. (c) Simplified model of the recoil of the solid particle in an optical trap approximated as a 1D harmonic oscillator.   %\emph{???The recoil happens at the moment, when alpha particle leaves the solid particle, if daughter atom has already stopped in the solid particle???}.} 
	}
	\label{fig:recoil}
\end{figure}

Nuclear decay is a fundamental process by which an unstable atom decays into a daughter atom and emits radiation. This radiation can be carried by $\gamma-$photons or by $\alpha-$ and $\beta-$particles. % with mass equal to four times of a proton mass and an electron mass, respectively. %This process is described by quantum mechanics, and rely on the fact that there is a non-zero probability of a nuclear particle tunneling through the nucleus barrier.
We begin by considering a nuclear decay occurring in an atom of some radioactive material contained within a small solid particle in free space (Fig. \ref{fig:recoil} (a), (b)). Accurate detection and characterization of the recoil through kinematics of the solid particle is possible if the full linear momentum of the daughter atom ($\bf{p_{d}}$) and none of the linear momentum of nuclear particle ($\bf{p_{n}}$) are transferred to the solid particle ($\bf{p_{s}}$). This happens when the solid particle size is such that the nuclear particle ($\alpha$/$\beta$/$\gamma$) escapes the solid particle without any interactions, while the daughter atom remains within the solid. For example, for $^{238}$PuO$_2$ this condition is satisfied for particle diameters between 10~nm and 10~$\mu$m \cite{PseudoEvaporation}. According to the linear momentum conservation law, the total linear momentum of the system before the nuclear decay (parent atom and solid particle) is equal to its counterpart after the decay (daughter, nuclear particle, and solid particle). Under the assumption that the solid particle and the parent atom are at rest before the decay, and the daughter remains within the solid particle after the decay, we find:  
\begin{equation}\label{Momentum_Cons}
%&\boldsymbol{p_{n}}+\boldsymbol{p_{s}}=0\;,\\
\boldsymbol{p_{s}}=-\boldsymbol{p_{n}}\;\;\;\;.
\end{equation}
Distinct decay types ($\alpha$/$\beta$/$\gamma$) result in different dependences of the linear momentum $\boldsymbol{p_{n}}$ (and consequently $\boldsymbol{p_{s}}$ through Eq.~\ref{Momentum_Cons}) on kinetic energy ($E_{kin}$) and mass of the emitted nuclear particle, and the proximity to the relativistic regime. First, we consider the following examples of alpha decay:
%\emph{Alpha decay.}
\begin{equation}\label{Eq-Pu239}
^{239}\mbox{Pu}\rightarrow\;^{235}\mbox{U}^{2-}+\alpha^{2+},
\end{equation}
where the $^{239}$Pu half-life is 24110 years, and also,
\begin{equation}
^{235}\mbox{U}\rightarrow\;^{231}\mbox{Th}^{2-}+\alpha^{2+},
\end{equation}
where the $^{235}$U half-life is 703.8 million years. The kinetic energy ($E_{kin}$) of an $\alpha-$particle in the nuclear decay process in Eq. (\ref{Eq-Pu239}) is 5.16~MeV~\cite{NuclearDataTables}, which is significantly less than its rest mass energy: $E_{0_\alpha}=m_\alpha c^2\approx3.7$~GeV, where $c$ is the speed of light in vacuum and $m_\alpha$ is the mass of the $\alpha-$particle. Therefore, for this non-relativistic case, the momentum of the $\alpha-$particle is:
\begin{equation}\label{eq:p_alpha}
p_{\alpha}= m_\alpha V_\alpha= \sqrt{2m_\alpha E_{kin_\alpha}}
\end{equation}

%\emph{Beta decay.} 
Another common type of nuclear decay is $\beta-$decay. In comparison with $\alpha-$decay, where an $\alpha-$particle is present in the nucleus before the decay takes place, and it is emitted with a unique kinetic energy, a $\beta-$particle and neutrino are simultaneously born in the process of the decay, and have a continuous energy spectrum. Let us consider the following example: 
\begin{equation}\label{eq:beta_Pu241}
^{241}\mbox{Pu}\rightarrow\;^{241}\mbox{Am}^{+}+\beta^{-}+\HepAntiParticle{\nu}{e}{},
\end{equation}
where $\HepAntiParticle{\nu}{e}{}$ is an electron-type antineutrino and the $^{241}$Pu half-life is 14.3 years. The mean and maximum kinetic energies of the $\beta-$particle for this example are 5.2~keV and 20.8~keV, respectively~\cite{NuclearDataTables}. The typical kinetic energy of a $\beta-$particle created during the decay of different elements vary in the range from a few keVs to a few MeVs, while its rest mass energy $E_{0_e}=m_e c^2=0.511$~MeV, where $m_e$ is the mass of the electron. For the general relativistic case, the momenta of the $\beta-$particle and the neutrino are:
\begin{align}\label{eq:p_beta}
\begin{split}
p_{\beta}&=\gamma \beta m_e c=m_e c\sqrt{\gamma_e^2-1},\\
p_{\nu}&=m_\nu c\sqrt{\gamma_\nu^2-1}
\end{split}
\end{align}
where $\gamma_{e,\nu}=1+\frac{E_{kin_{e,\nu}}}{E_{0_{e,\nu}}}$ are the Lorentz factors for the electron and neutrino, respectively.
It is important to note that the neutrino rest mass is very small: the upper limit of the effective Majorana neutrino mass varies in the range 0.061 - 0.165~eV \cite{Majorana}. However, the kinetic energy of the neutrino can reach up to a few MeVs, and it exhibits a continuous momentum spectrum correlated with the corresponding $\beta-$particle counterpart~\cite{BetaSpectra_Sherwin1,BetaSpectra_Sherwin2}.   

%\emph{Gamma decay.} 
Finally, we consider the case of gamma decay, the process in which an atomic nucleus emits a $\gamma-$photon while transitioning from the excited state to a lower excited or the ground state. For instance, this can happen when $\alpha-$ or $\beta-$decay yield a daughter nucleus to be in an excited state. For example, the $\alpha-$decay of $^{241}$Am :
\begin{equation}\label{alpha_for_gamma}
^{241}\mbox{Am}\rightarrow\;^{237}\mbox{Np}^{2-}+\alpha^{2+},
\end{equation}
yields a ground state $^{237}$Np daughter with probability of only $0.37\%$, while most probably ($84.8\%$), it yields a daughter in an excited state at $59.54$~keV~\cite{NuclearDataTables}. This $\alpha-$decay is followed by the $\gamma-$decay almost immediately (67~ns)~\cite{NuclearDataTables}, which results in an additional momentum kick to the solid particle in a random direction. This is under the assumption that the $\gamma-$photon leaves the solid particle, which is a very likely case for micron-sized particles. The interaction of gamma-rays with matter strongly depends on their energy and the material properties, so we leave the probability estimates for that process beyond the scope of this paper, while more information can be found in the literature~\cite{Gamma-matter_LANL}. The kinetic energies ($E_{kin_\gamma}=\hbar w_\gamma$) for a massless $\gamma-$particle are typically in the range  $10$~keV - $4$~MeV~\cite{Nuclear_Book} and are related to its momentum via the linear dispersion relation:
\begin{equation}\label{eq:p_gamma}
 p_{\gamma}=\hbar k_{\gamma}=E_{kin_{\gamma}}/c. 
\end{equation}
For the case discussed in Eq.~(\ref{alpha_for_gamma}), our estimates show that the recoil momentum caused by the gamma decay from the most probable excited state of $^{237}$Np ($59.54$~keV) to the ground state is $\sim3.5$ orders of magnitude smaller than the preceding alpha recoil momentum. Therefore, for  most cases, gamma recoil is negligible when compared to the recoil from $\alpha-$decay and, hence, can be ignored for practical applications (see sec. \ref{spectroscopy}).
In contrast, the recoil due to $\beta-$decay and that of the associated $\gamma-$decay can be of the same order. For instance, beta decay in $^{238}\mbox{Np}$~\cite{Np-238}:
\begin{equation}\label{beta_for_gamma}
^{238}\mbox{Np}\rightarrow\;^{238}\mbox{Pu}^{+}+\beta^{-}+\HepAntiParticle{\nu}{e}{},
\end{equation}
results in $^{238}\mbox{Pu}$ being in two excited states: $E_1=44.051$~keV and $E_2=1028.542$~keV, with probabilities of $41.1\%$ and $44.8\%$, respectively \cite{NuclearDataTables,Pu-238}.  For simplicity, we assume that the neutrino momentum is zero, and the beta particle possesses maximum kinetic energies of $1248$~keV and $263$~keV, for $E_1$ and $E_2$, respectively~\cite{NuclearDataTables}. For the first excited level ($44.051$~keV), we find the ratio of recoil momenta between $\beta-$ and $\gamma-$decays to be $\sim38$. The half-life time of $^{238}\mbox{Pu}$ on this level is $\sim175$~ps~\cite{NuclearDataTables}, which means that the randomly directed $\gamma-$kick will cause spectrum broadening of the $\beta-$recoil in this simplified neutrino-less approximation. For the most probable excited level ($1028.542$~keV) of $^{238}\mbox{Pu}$, our estimates show that the $\gamma-$kick is actually $\sim1.6$ times larger than the biggest possible preceding kick from $\beta-$decay. To the best of our knowledge, the half-life time at this level has not been measured yet, but if it is less than several $\mu s$, it will significantly complicate beta decay characterization in this particular example. Quantifying $\beta-$decay in a solid particle with our method is additionally complicated because of the spectrum broadening due to neutrino emission, so we suppose that the proposed technique is limited to samples where $\gamma-$decay is significantly postponed from $\beta-$decay, or is not present at all (see sec. \ref{spectroscopy}).

\begin{figure}[t]
	\includegraphics[width=0.49\textwidth]{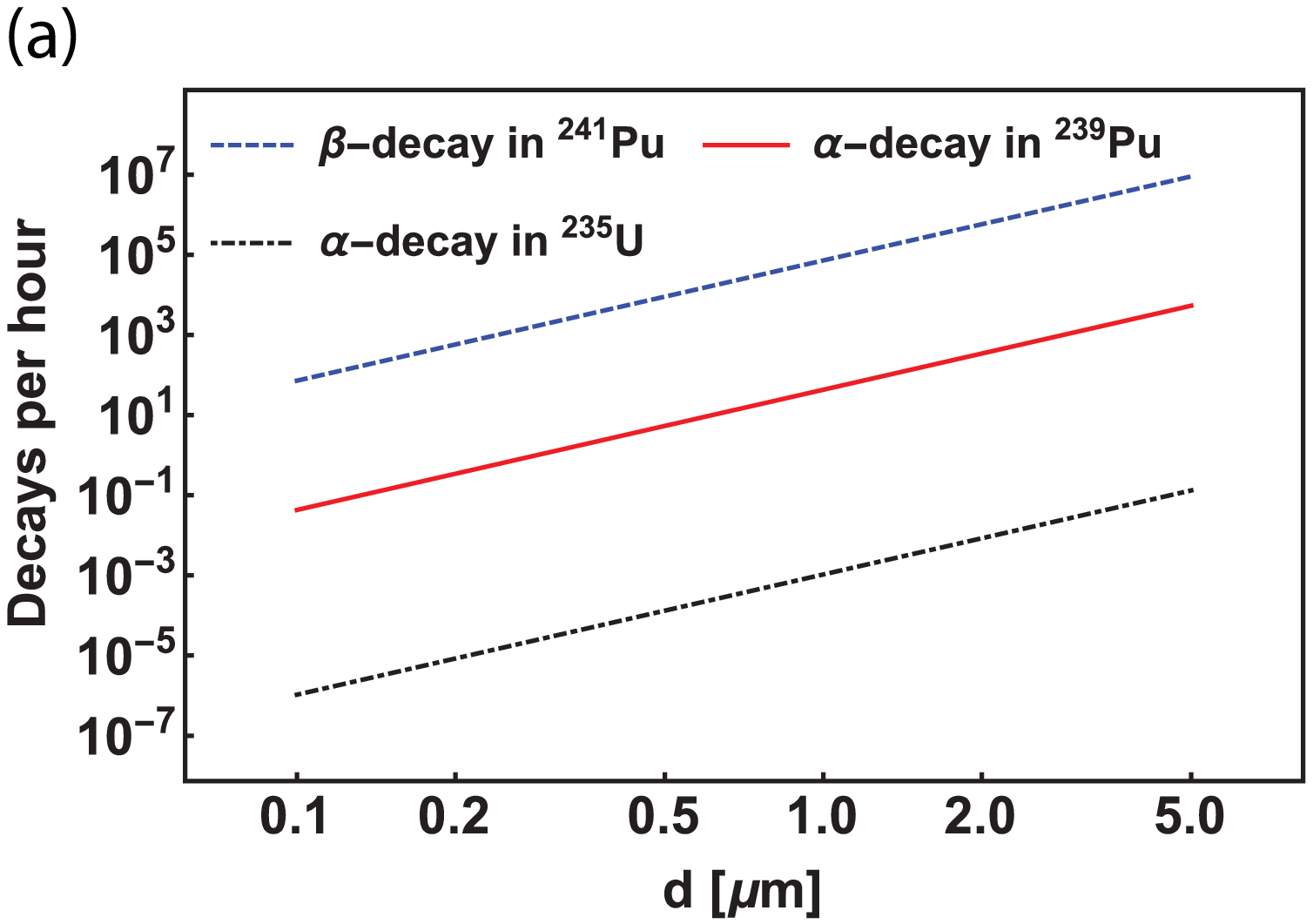}
	\includegraphics[width=0.49\textwidth]{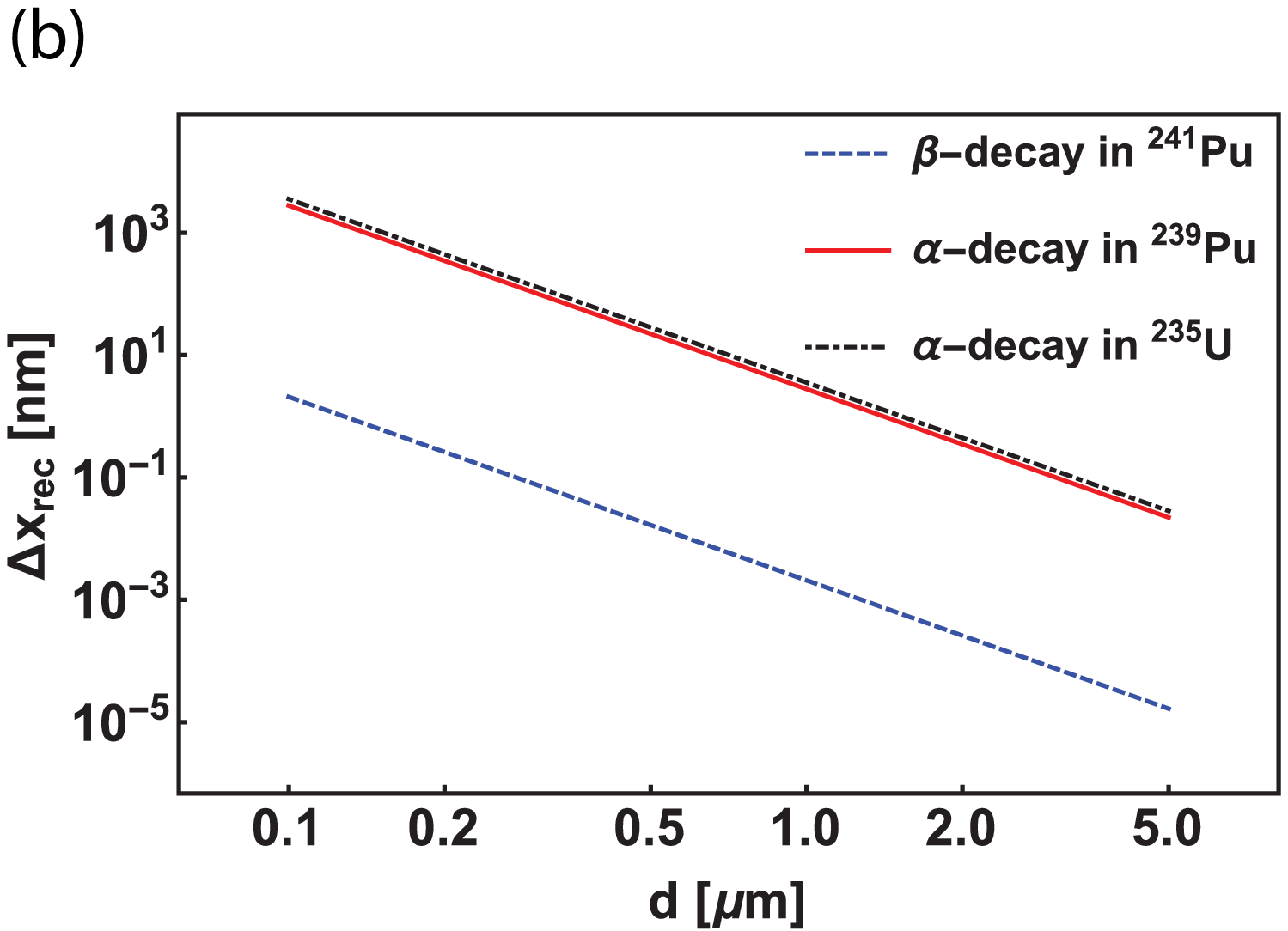}
	\caption{(a) Recoil rate as a function of particle diameter for $^{241}$Pu, $^{239}$Pu, and $^{235}$U particles in oxide form (PuO${_2}$ and UO${_2}$). (b) Particle displacement due to nuclear recoil as a function of particle diameter in a 1 kHz optical trap. The $\beta-$recoil is calculated under the assumption of a zero neutrino momentum and mean $\beta-$energy.} 
	\label{fig:rate}
\end{figure}

Due to the small size of the particles considered here, it is essential to estimate the rate at which nuclear decay happens, and how it scales with the size of the particle. 
Nuclear exponential decay is described by the equation:
\begin{equation}\label{exp_decay}
N(t)=N_0 e^{-t/\tau},
\end{equation}
where $N_0$ is the initial number of radioactive atoms, $N(t)$ is the number of not decayed atoms, $\tau=t_{1/2}/\ln(2)$ and $t_{1/2}$ are the mean lifetime and the half-life of the decaying atoms, respectively~\cite{Nuclear_Book}.
The number of nuclear decays per time interval from $t_1$ to $t_1+\Delta t$ can be found as:
\begin{equation}\label{decay_rate}
\Delta N=N(t_1)-N(t_1+\Delta t)\simeq N(t_1) \Delta t/\tau,
\end{equation}
if $\Delta t<<\tau$.
Therefore, the average recoil rate ($r=\Delta N/\Delta t$) is proportional to the number of radioactive atoms, and inversely proportional to their half-life. In addition, the rate of recoil depends on the isotopic composition of the sample and the radioactivity of each isotope present.
For simplicity, consider a solid particle of a radioactive element ($R$) oxide, $R_AO_B$, approximated by a sphere with diameter, $d$. Then, the mass of this particle is: $m=\frac{1}{6}\pi\rho d^3$, where $\rho$ is the average particle density. Hence, the number of radioactive atoms in this particle is: $N_{R}=\frac{m}{M}N_aA$, where $N_a$ is  Avogadro's number, and $M$ is the molar mass of the oxide. For example, as presented in Figure \ref{fig:rate} (a), for a 4.5 $\mu$m $U_3O_8$ particle with $100\%$ $^{235}$U isotope abundance (half-life of 703.8 million years) we find $r=\frac{N_U\ln{2}}{t_{1/2}}=2$ $\alpha-$decays per day.  In comparison, for $^{241}$Pu (half-life of 14.3 years) in a particle of PuO$_2$ thirty times smaller ($d=150$ nm), we expect $\sim4$ $\beta-$decays per minute.  

In order to observe a statistically significant number of recoils, it is essential to localize a particle in space. One of the most advanced methods to achieve this localization is via optical trapping, which allows the levitation of a particle by one or several laser beams \cite{Ashkin,review}. The motion of a solid particle in an optical trap near the point-of-rest is well-described by a harmonic oscillator. Using a 1D approximation, we find that the displacement of a particle initially at rest due to the nuclear recoil is:   
\begin{equation}\label{eq:x_recoil}
\Delta x_{rec}=\frac{V_{s}}{\omega}=\frac{p_{s}}{m_{s}\omega}=\frac{p_{n}}{m_{s}\omega}\;\;\;,
\end{equation}
where $V_s$ and $m_s$ are the velocity and mass of the solid particle, and $\omega$ is the trap frequency.
As shown above, the displacement caused by nuclear recoil is inversely proportional to the particle mass and, hence, it scales with the diameter of the particle as $\sim d^{-3}$ (Fig. \ref{fig:rate} (b)). For example, for a 100 nm particle of pure $^{235}U_3O_8$ undergoing alpha recoil ($E_{{kin}}\sim4.5$ MeV)~\cite{NuclearDataTables} in a trap with $\omega=1$ kHz, the expected displacement is $\sim 1$ $\mu$m. This displacement is well above the positional resolution of recently demonstrated  experimental techniques, which are capable of detecting displacements of several picometers\ \cite{T_Li, Novotny}. 

Levitating particles of oxides of actinides such as uranium, plutonium, and americium, should be feasible with conventional gradient force traps~\cite{Ashkin,review}, since these compounds exhibit dielectric optical properties. In contrast, trapping particles of pure metal actinides may be problematic because they exhibit a high degree of optical absorption diminishing the gradient force and as a result preventing the levitation in high vacuum. An exception to this complication for metallic particles can be made when the diameter is much smaller than the trapping laser wavelength~\cite{Gold-nanoparticles1,Gold-nanoparticles2}, since the skin-depth of metals is typically of the order of at least several wavelengths in the optical frequency range. Photon absorption can dramatically increase the internal temperature of the particle, even up to the melting point of the material, as has been shown for silicon oxide particles, for example~\cite{Hebestreit, Nano}. Indeed, practically the only cooling mechanism of the trapped particle in high vacuum is black body radiation, which has been demonstrated to be fairly inefficient~\cite{Hebestreit, Nano}. However, the melting temperatures of actinides are much higher than those of silica or metals, thus mitigating the issue. While the optical properties of materials may impose limitations on the type of samples that can be investigated with the proposed method, we emphasize that actinide oxides, the most common of the samples of interest for nuclear forensics, can be effectively trapped and studied.

In principle, nuclear decay can happen at any location within the solid particle. If it does not happen at the center-of-mass (COM) position, the solid particle will eventually start rotating due to the conservation of the total angular momentum of the system (the total external torque acting on the system is zero). Before decay takes place, the angular momenta of the decaying atom at rest ($\boldsymbol{L_a}=\boldsymbol{r_a}\times\boldsymbol{p_a}=0$) and the non-rotating solid particle ($\boldsymbol{L_s}$=0) result in a total zero angular momentum of the system. Immediately after an off-center decay, the angular momentum of the solid particle still remains zero, while the angular momentum of the nuclear particle $\boldsymbol{L_n}=\boldsymbol{r_a}\times\boldsymbol{p_n}$ and the daughter atom $\boldsymbol{L_d}=\boldsymbol{r_a}\times\boldsymbol{p_d}$ (both are considered here with respect to the origin of the coordinate system, placed at the COM of the solid particle for convenience) must compensate each other to provide with a total zero angular momentum, since it is conserved. This immediately results in $-\boldsymbol{p_n}=\boldsymbol{p_d}$, which can be independently derived from the linear momentum conservation principle right after the decay happens (the total external force acting on the system is zero). The nuclear particle, which we assume leaves the solid particle without any interaction, can be excluded from further consideration without loss of generality. Both the total linear momentum and the angular momentum of the two-body system are conserved during the interaction of the daughter and the solid particle. These quantities are equal to the corresponding values for the daughter right after decay: $\boldsymbol{p_d}$ and $\boldsymbol{L_d}$. Assuming that the daughter eventually stops within the solid, \emph{e.g.} becomes embedded as part of it, such a solid particle will continue its motion, characterized by the linear momentum $\boldsymbol{p_{s}}=-\boldsymbol{p_n}$ and angular momentum $\boldsymbol{L_{s}}=-\boldsymbol{r_a}\times\boldsymbol{p_n}$. The non-zero angular momentum ($\boldsymbol{L_{s}}$) characterizes the rotation of the solid particle around its COM (the mass of the daughter is negligible compared to that of the solid particle), but the linear momentum ($\boldsymbol{p_{s}}$) solely characterizes the translational motion of the solid particle. Thus, the resulting linear momentum of the solid particle is independent of the particular location where the decay occurs. Therefore, by measuring the displacement of the solid particle with known mass ($m_{s}$) in an optical trap with specific frequency ($\omega$), we can reconstruct the linear momentum (through Eq.~\ref{eq:x_recoil}) and the kinetic energy (through Equations~(\ref{eq:p_alpha}, \ref{eq:p_beta}, or~\ref{eq:p_gamma})) of the nuclear particle emitted in the process of radioactive decay. 

\section{Practical and fundamental limits for resolving nuclear recoil displacements}\label{limits}

\begin{figure}[t]
	\includegraphics[width=0.49\textwidth]{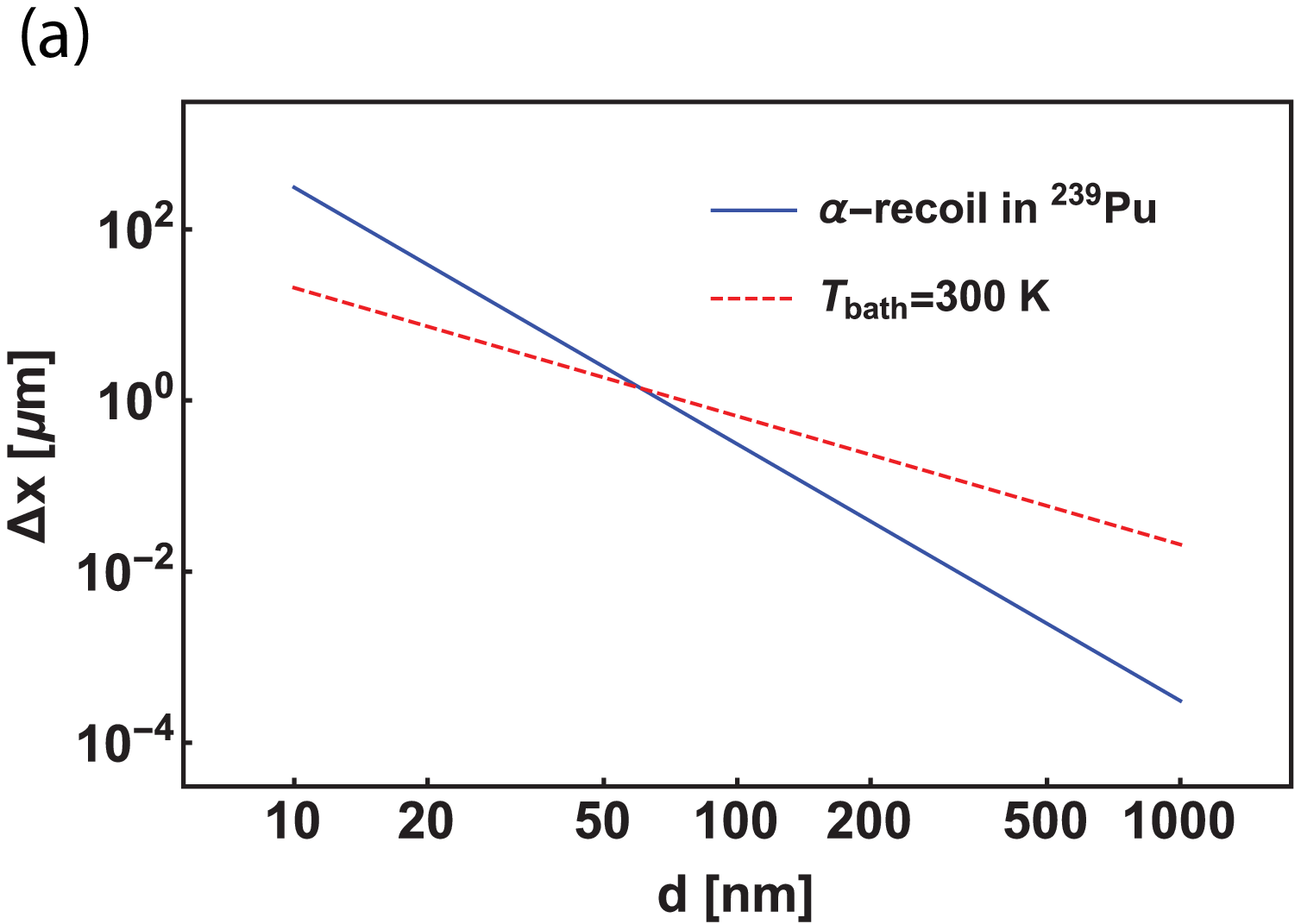}
	\includegraphics[width=0.49\textwidth]{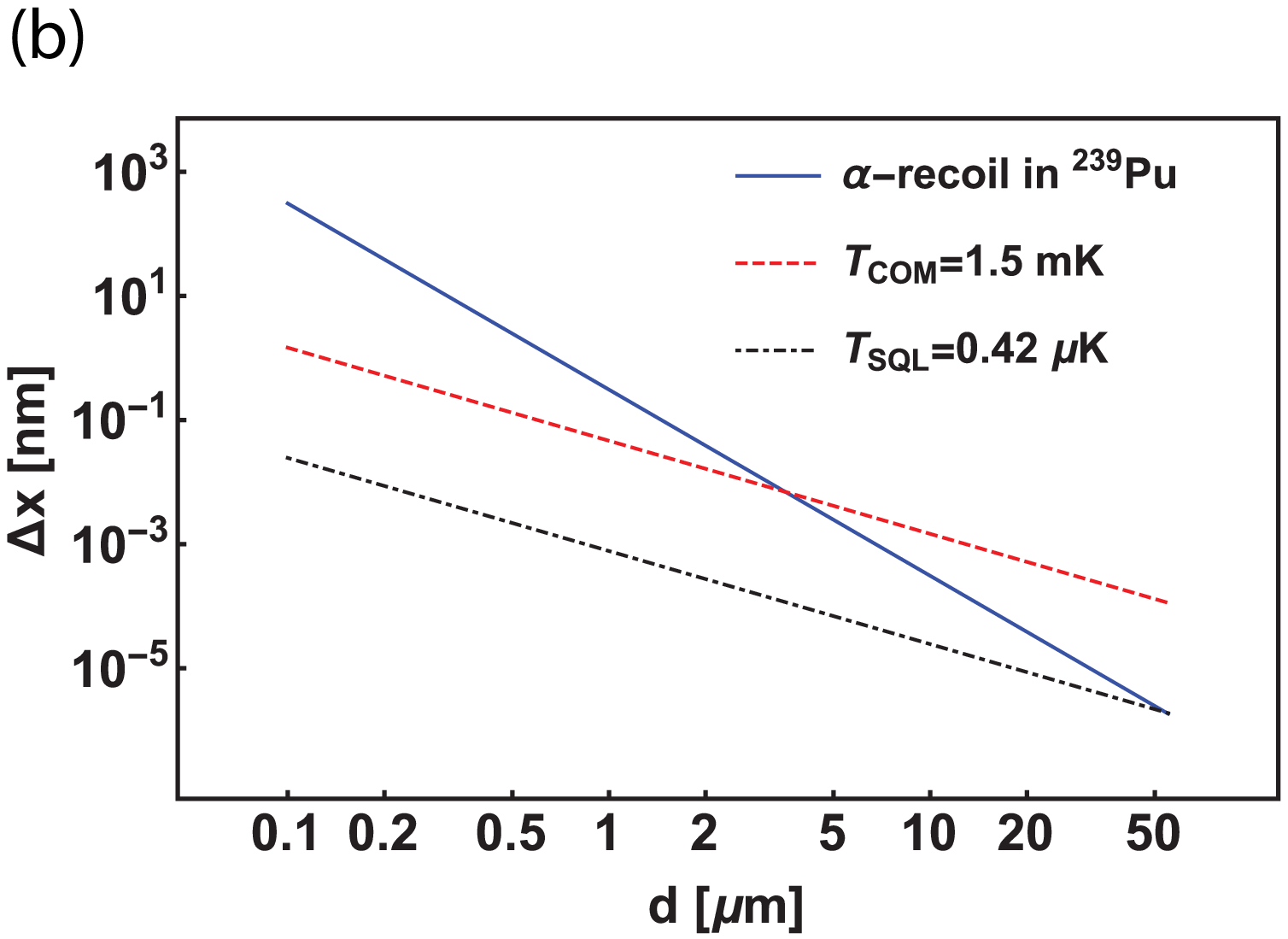}
	\caption{Performance of a 9 kHz optical trap: (a) The displacement due to $\alpha-$decay in a $^{239}$Pu sample is compared to the thermal noise at ambient conditions. (b) The displacement due to $\alpha-$decay in a $^{239}$Pu sample is compared to the amplitude of the COM motion of a particle cooled to 1.5 mK,  and also compared to the motion of a particle cooled to the $T_{SQL}$ of 0.42 $\mu$K, rescaled from $T_{SQL}\sim5.6$ $\mu$K for a 120 kHz trap, as reported in \cite{Novotny, Novotny+}.}
	\label{Recoil_limits}
\end{figure}

A trapped particle interacts with its surrounding gas via collisions. At thermal equilibrium, the average energy of the COM motion in the trap is $\sim k_BT$, where $T$ is the temperature of the surrounding gas and $k_B$ is the Boltzmann constant. The Brownian motion of the trapped particle plays the role of positional noise with amplitude: $\Delta x_{bath}=\sqrt{\frac{k_B T}{m_{s}\omega^2}}$. For a $^{239}$PuO$_2$ particle, Figure \ref{Recoil_limits} (a) illustrates how this noise limits the detection of recoils for large particles. Note that for particles less than 60~nm, recoils can be detected due to the following reason. While both the displacements due to nuclear recoil as well as the displacements due to collisions with the background gas depend equally on the frequency of the trap, they scale differently with the particle mass, $1/m_{s}$ and $1/\sqrt{m_{s}}$, respectively. The observation of the recoil for such small particles requires a long acquisition time (see Fig.~\ref{fig:rate} (a)), unless one targets isotopes of high radioactivity.% for example: $^{208}$Po ($t_{1/2}\sim$ 2.9 years), $^{210}$Po ($t_{1/2}\sim$ 138 days). 

Since thermal noise obscures nuclear recoils for large particles, it is essential to isolate such particles from the environment by removing surrounding air to high vacuum levels, and by cooling their COM motion via positional feedback \cite{T_Li, Novotny}. Active feedback cooling can be realized, for example, by applying additional radiation pressure to the trapping beam in order to compensate against particle displacements caused by collisions with the residual gas~\cite{T_Li}. Using this technique, a temperature of $\sim1.5$ mK has been reached for a 3~$\mu$m silica micro-sphere at a trapping frequency of $\sim$ 9~kHz along one of the axes, at a residual pressure of $\sim5\cdot10^{-5}$~mbar~\cite{T_Li}. It has also been reported that a particle COM temperature of $\sim0.1$~mK can be reached with a demonstrated sensitivity of the detection system of 39~fm~Hz$^{-1/2}$, upon improvement of the cooling technique~\cite{T_Li}. In fact, this technique is limited in practice by the resolution of the detector used to measure the particle displacement and its instantaneous velocity, and fundamentally by the photon recoil heating from the feedback laser beams. In the Rayleigh regime, when the particle size is smaller than the wavelength of the trapping laser light, the photon recoil limit can be defeated by applying a parametric feedback technique, a different method of feedback cooling~\cite{Novotny}. A COM temperature of roughly $450\;\mu$K has been demonstrated for 100-nm fused silica particles by using the parametric feedback cooling method~\cite{Novotny1}. This approach works only for nanoparticles, since it relies on treating the particle as a phase-coherent antennae in the near field regime.  

Finally, the standard quantum limit (SQL) defines the fundamental cooling limit in an optical trap. This limit is achieved when the uncertainty of the scattered photon momentum is equal to the uncertainty of the measured particle position. If the SQL is reached, the position accuracy scales as $\Delta x\sim\sqrt{\hbar/(m_{s}\omega)}$ and it is independent from the laser power directly~\cite{Novotny+}. However, the SQL is reached for a particular photon number which depends on the particle size and the frequency of the trap. Therefore, the laser power and the trap geometry have to be properly adjusted for a given particle to reach the SQL. The SQL temperature itself does not depend on particle mass or size and scales linearly with the trap frequency: $T_{SQL}\sim\hbar\omega/k_B{\tiny}$. The fundamental limit sets the lowest achievable temperature of $\sim5.6$~$\mu$K at a trapping frequency of 120 kHz~\cite{Novotny,Novotny+}. Rescaling the SQL temperature to a 9 kHz trap yields $T_{SQL}\sim0.42$~$\mu$K. Figure~\ref{Recoil_limits}~(b) compares a typical alpha recoil displacement, the average particle motion at the experimentally demonstrated temperature of 1.5~mK~\cite{T_Li, Comment}, and the motion at the fundamental limit, as a function of  $^{239}$Pu particle diameter in a 9 kHz trap. This trap frequency can be reached for particles over a wide size range by adjusting the laser parameters and trap configuration. The intersection of the recoil curve with the experimental curve demonstrates that single recoil detection is possible for particles smaller than $3$~$\mu$m at experimentally demonstrated temperatures.
Moreover, the intersection of the recoil curve with the SQL curve at $d\sim50$~$\mu$m, suggests that a single $\alpha-$recoil can be resolved for practically any trappable particle upon reaching the SQL. Theoretical estimates~\cite{SCat, Zoller, SQL1} demonstrate that COM temperatures close to the quantum-mechanical ground state can be reached experimentally, and many research groups are working towards this limit, motivated by studies in non-Newtonian gravitation, Casimir force sensing, measuring vacuum friction, etc.~\cite{T_Li, Novotny, Geraci1, Verhagen, CF,Gieseler1}. Finally, for particles in the range 100~-~500~nm, the recoil displacement is several orders of magnitude larger than the experimentally demonstrated positional noise. This gap is even larger at SQL temperatures. This suggests that a single recoil can not only be detected, but the kinetic energy of the nuclear particle can be reconstructed with good accuracy using the formalism described in Section~\ref{TM}, opening up the possibility for resolving the decays of individual isotopes.

%\textsl{}\iffalse 

\section{Nuclear recoil spectroscopy}\label{spectroscopy}

As discussed above, efficient cooling of the COM motion of a particle decreases its displacements due to Brownian motion to significantly smaller amplitudes than the displacement due to nuclear recoil. However, the direct instantaneous observation of individual recoil events while the particle is under positional feedback can be obscured, because the feedback will suppress the displacement due to recoil. Yet, the information about the recoil causing a linear momentum increase would be present in the feedback signal associated with the particle motion. In particular, if the displacement due to nuclear recoil is larger than the resolution of the positional detection system, each event will cause a phase shift of the signal and increase its oscillation amplitude. An alternative approach is to switch the feedback off as soon as the particle has been cooled to the lowest temperature, followed by switching the feedback on once the COM temperature is too high to observe the recoil. In the absence of feedback, heating will cause the noise amplitude to grow. Generally speaking, the heating rate is due to collisions with gas molecules $\Gamma_{gas}$, photon recoils $\Gamma_{photon}$, and other experimental noise $\Gamma_{exp}$, such as trapping laser noise and mechanical vibrations. In practice, $\Gamma_{exp}$ dominates the heating rate, providing us with a time window of the order of a few seconds to observe the recoil ($\Gamma_{exp}\approx200$~mHz, as experimentally demonstrated in reference~\cite{Geraci} for a 300-nm fused-silica sphere at $5\cdot10^{-6}$~Torr). Since the time it takes to cool the particle is typically much shorter ($\sim$10~ms), the duty cycle during which one can observe nuclear recoils is expected to be above 90\% for such pulsed sequence. Ultimately, there is a trade off between the duty cycle and the affordable noise level.  
Since the COM temperature depends linearly on the interaction time at pressures of the order of  $\sim10^{-5}$ mbar~\cite{Hebestreit}, this results in a shorter time interval when the feedback is turned off. In contrast, the parametric feedback damping rate depends quadratically on the positional noise amplitude. Therefore, the effective time to cool down the motion of the particle is practically independent of the upper temperature~\cite{Hebestreit}. In other words, in order to decrease the positional noise level, and hence increase the resolution of a recoil displacement measurement, one may sacrifice the duty cycle and prolong the observation time.

\begin{figure}[t]
	\includegraphics[width=0.49\textwidth]{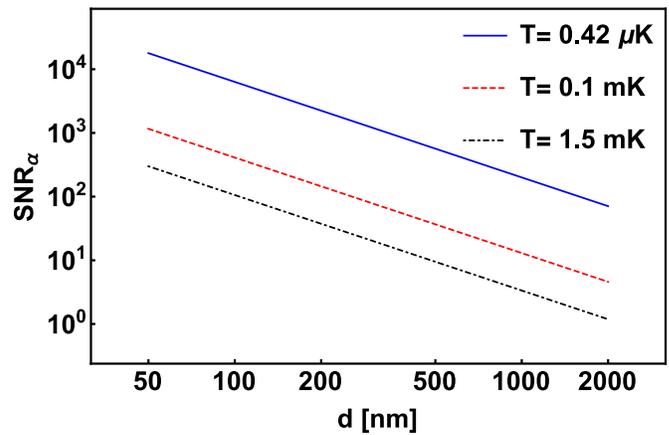}
	\caption{SNR for $\alpha-$recoil spectroscopy in a 9~kHz optical trap for a $^{239}$PuO$_2$ sample at different temperatures: $T_{COM}=1.5$~mK, $T_{noise}=0.1$~mK (as demonstrated in~\cite{T_Li}), and $T_{SQL}=0.42~\mu$K.}
	\label{fig:SNR}
\end{figure}

Measurement of a nuclear recoil with good positional resolution allows resolving the kinetic energies of the emitted decay particles. As mentioned in section~\ref{limits}, the COM particle motion plays the role of positional noise for recoil detection. In the regime where particle oscillations in the trap are stochastic, the recoil displacement adds to the instantaneous particle motion. Even in the regime when the particle motion is coherent, the decay-initiated recoil still happens with a random phase with respect to the particle oscillation in the trap. Nevertheless, assuming randomly distributed Gaussian noise on the measured recoil displacement, we introduce the signal-to-noise ratio (SNR) for the recoil displacement along $x-$axis as SNR$_{x}=\Delta x_{rec}/\Delta x_{bath}$. For alpha decay, the kinetic energy of the $\alpha-$particle scales as $\sim\Delta x_{rec}^2$ according to Eqs.~(\ref{eq:p_alpha}) and Eq.~(\ref{eq:x_recoil}). For a SNR$_{x}>>1$, using the Taylor series expansion and keeping only linear terms, we find: SNR$_{\alpha}=E_{rec}/\Delta E\simeq 0.5\cdot \Delta x_{rec}/\Delta x_{bath}\sim m_{s}^{-1/2}\sim \rho^{-1/2}\cdot d^{-3/2}$. Figure~\ref{fig:SNR} shows the SNR$_{\alpha}$ as a function of  diameter for $\alpha-$decay within a  $^{239}$PuO$_2$ particle, as described in Eq.~(\ref{Eq-Pu239}). The dependence is presented for this particle in a 9~kHz optical trap at different temperatures related to the fundamental limit, $T_{SQL}=0.42~\mu$K, the noise of the position detection system, $T_{noise}=0.1$~mK as demonstrated in~\cite{T_Li}, and the temperature of the center-of-mass motion, $T_{COM}=1.5$~mK, achievable with one of the two previously described cooling methods depending on the particle size~\cite{T_Li,Novotny1} (see section~\ref{limits} for details). Figure~\ref{fig:SNR} shows that cooling the motion of the particle to lower temperatures allows to reach higher SNRs, expanding the size range towards larger particles, and hence, allowing the detection of recoil in less radioactive materials. In contrast, at higher temperatures, detecting $\alpha-$recoil with good resolution is limited to smaller particles, which requires longer acquisition times.

\begin{figure}[t]
	\includegraphics[width=0.50\textwidth]{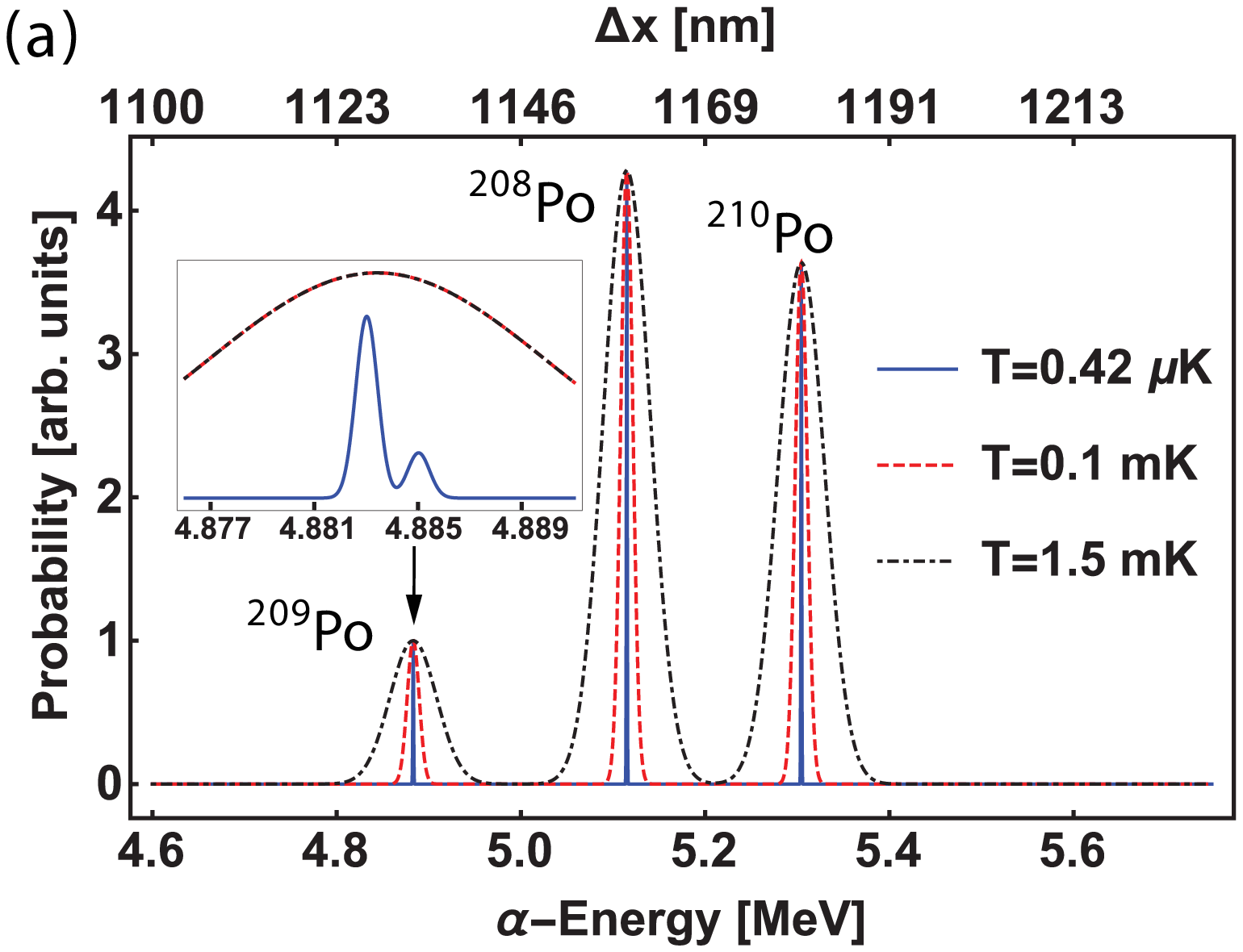}
	\\
	\includegraphics[width=0.49\textwidth]{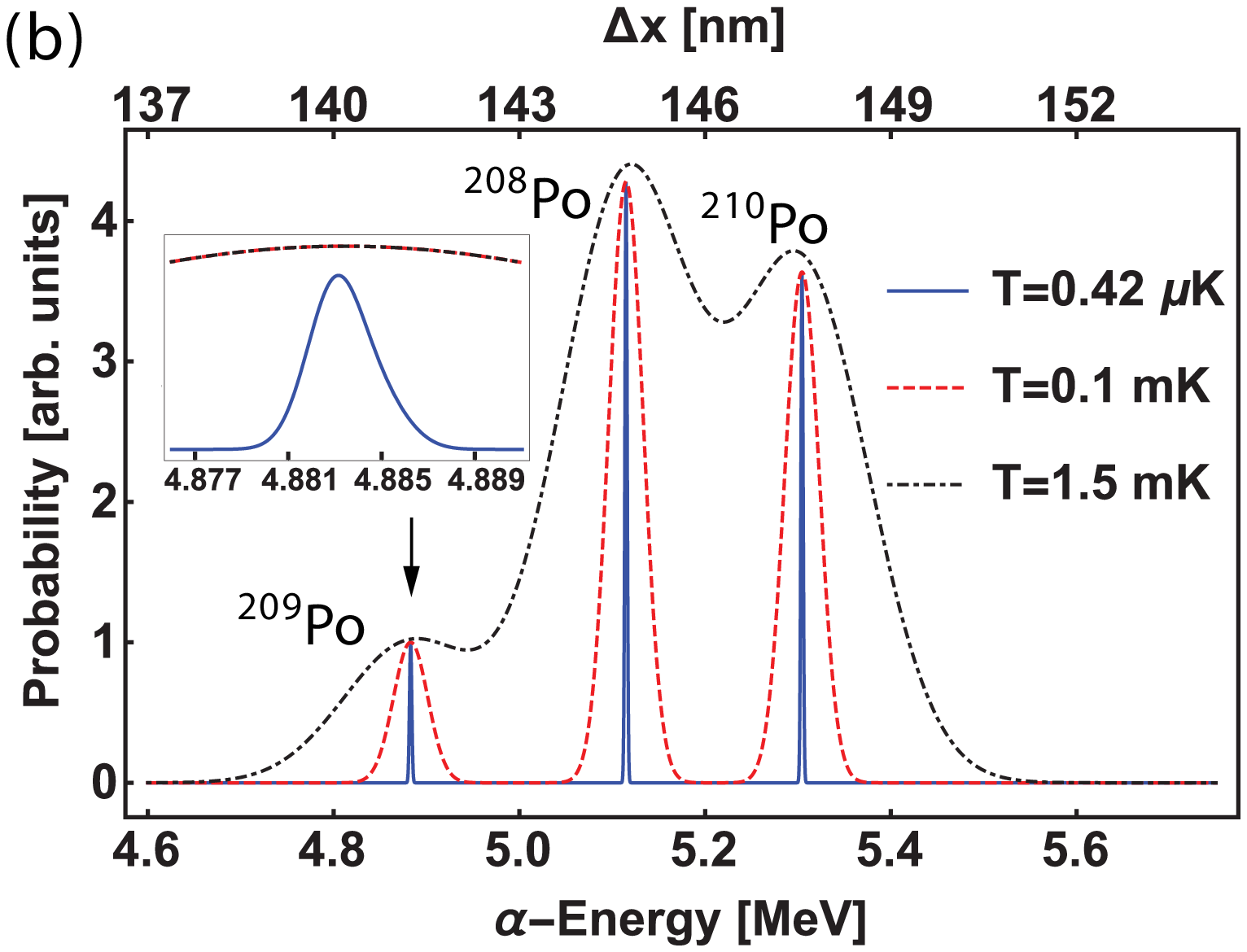}
	\caption{$\alpha-$recoil spectroscopy in a 9~kHz optical trap for a $^{209}$Po sample containing 9\% $^{208}$Po and 1\% $^{210}$Po: (a) $d=70$~nm and (b) $d=140$~nm. The insets demonstrate the resolved (a) and unresolved (b) nuclear energy structure of $^{209}$Po decay: $\alpha-$particle kinetic energies of $\sim4883$ keV and $\sim4885$ keV are related to the $^{205}$Pb daughter being in the excited and ground states, respectively. The upper horizontal axes relate the displacement of the solid particles to the kinetic energy of the emitted nuclear particles.}
	\label{fig:alpha_Po}
\end{figure}

To ensure adequate acquisition times at experimentally demonstrated temperatures, consider a highly radioactive sample of $^{209}$Po containing 9\% $^{208}$Po and 1\% $^{210}$Po. The half-lives of these isotopes are 124~years, 2.9~years and 138~days, respectively. Polonium isotopes undergo $\alpha-$decay to Pb isotopes  according to the equations:
\begin{align}
\begin{split}\label{Eq:alpha_Po}
^{208}\mbox{Po}\rightarrow\;^{204}\mbox{Pb}^{2-}+\alpha^{2+},\\
^{209}\mbox{Po}\rightarrow\;^{205}\mbox{Pb}^{2-}+\alpha^{2+},\\
^{210}\mbox{Po}\rightarrow\;^{206}\mbox{Pb}^{2-}+\alpha^{2+}.
\end{split}
\end{align}
$^{209}\mbox{Po}$ decays predominantly to the excited level (2.3~keV) and ground level of $^{205}\mbox{Pb}$ with probabilities of  79.2~\% and 19.7~\%, respectively, while both $^{208}\mbox{Po}$ and $^{210}\mbox{Po}$ decay predominantly to the ground levels of their daughters~\cite{NuclearDataTables}. 
Figure~\ref{fig:alpha_Po} shows the $\alpha-$energy spectrum, constructed from calculated particle recoils in a 9~kHz trap at different temperatures using the formalism presented thus far. At $1.5$~mK, the energy peaks for each isotope are well-resolved for a 70-nm particle (Fig.~\ref{fig:alpha_Po}~(a)), allowing the extraction of isotope ratios. This can be done by integrating the areas under the curves while accounting for the decay rate of each isotope. The average recoil rate is one $\alpha-$recoil per $\sim26.3$~min for the $^{209}$Po isotope with abundance of 90\% in the 70-nm particle. In contrast, a shorter average time is required ($\sim3.3$ min) to observe a single $\alpha-$recoil in a 140-nm particle. However, for a particle of this size at 1.5~mK, the peaks are not resolved (Fig.~\ref{fig:alpha_Po}~(b)). At the COM temperature of $0.1$~mK (the position detection noise demonstrated in~\cite{T_Li}), the peaks are well-resolved for both particle sizes. Finally, upon reaching the SQL temperature (0.42~$\mu$K), it is possible to resolve the nuclear energy structure of $^{209}$Po decay in the 70-nm particle, despite the alpha energies being only $\sim2$ keV apart from each other.

The half-life on the excited level of $^{205}$Pb (2.3~keV) is 24.2~$\mu$s~\cite{NuclearDataTables}. The $\gamma-$recoil displacement is $\sim3$ and $\sim10$ times smaller than the positional noise at $T_{SQL}$ for 70-nm and 140-nm particles, respectively, and therefore, we can ignore the effect of the $\gamma-$decay spectrum broadening in the example of Figure~\ref{fig:alpha_Po}. However, for very small particles and/or low trap frequencies, the $\gamma-$decay broadening can be a dominant effect upon reaching the SQL temperature.
Consider two close alpha particle energies: $E_1$ and $E_2=E_1+\Delta E$. Assuming that $\gamma-$decay happens immediately after $\alpha-$decay (on the time scale of the detection system), there could be spectrum broadening, as mentioned in section~\ref{TM}. The difference in $\alpha$-recoil displacements can be calculated using Eqs.~(\ref{eq:p_alpha}) and (\ref{eq:x_recoil}). After linearization ($\Delta E<<E_{1,2}$) one finds: $\Delta x_{\alpha_1}-\Delta x_{\alpha_2}\simeq\Delta E\sqrt{2m_\alpha/E_1}/(2m_{s}\omega)$. Assuming the $\gamma-$particle is emitted with energy $\sim\Delta E$, or simply that $\gamma-$decay in the daughter occurs between these two energy levels, we find the $\gamma-$recoil displacement using Eqs.~(\ref{eq:p_gamma}) and (\ref{eq:x_recoil}): $\Delta x_\gamma=\Delta E/(m_{s}\omega c)$. The ratio $(\Delta x_{\alpha_1}- \Delta x_{\alpha_2})/\Delta x_\gamma\simeq\sqrt{m_\alpha c^2/(2E_1)}>>1$, since the $\alpha-$particle rest energy is $\sim3.7$~GeV, while the typical kinetic energy of an $\alpha-$particle rarely exceeds 10~MeV. Therefore, the $\gamma-$recoil broadening would not be large enough to obscure the nuclear energy structure of the daughter atom.

\begin{figure}[t]
	\includegraphics[width=0.49\textwidth]{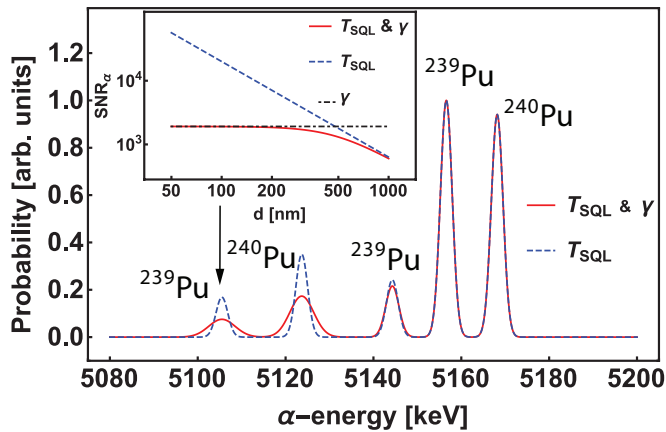}
	\caption{$\alpha-$recoil spectroscopy in a 0.9~kHz optical trap for an 80\% $^{239}$Pu/20\% $^{240}$Pu particle ($d=300$~nm) at $T_{SQL}=42~$nK with and without $\gamma-$recoil broadening. Inset demonstrates that the SNR for detecting $\alpha-$recoil ($E_\alpha=5105.5$~keV) in small particles is dominated by the associated $\gamma-$decay ($E_\gamma=51.7$~keV).}
	\label{fig:alpha_Pu239_240}
\end{figure}

Upon reaching the SQL temperature in a lower frequency trap, $\gamma-$recoil can play a significant role in broadening of the spectrum, since it scales as $(m\omega)^{-1}$, while the positional noise at $T_{SQL}$ scales as $(m\omega)^{-1/2}$. Consider an 80\% $^{239}$Pu/20\% $^{240}$Pu sample in a 0.9~kHz trap. According to Eq.~(\ref{Eq-Pu239}), $^{239}$Pu decays to the excited levels  of the daughter ($^{235}U$): $E_1=0.0765$~keV ($70.77\%$), $E_2=13.04$~keV ($17.11\%$),  $E_3=51.7$~keV ($11.94\%$), etc.~\cite{NuclearDataTables}. The half-life of these excited levels are 26~min, 500~ps, and 191~ps for $E_1$, $E_2$, and $E_3$, respectively~\cite{NuclearDataTables}. Similarly, $^{240}$Pu $\alpha-$decays to the ground ($72.8\%$) and excited, $E_1=45.2$~keV,  level ($27.1\%$) of the daughter ($^{236}U$), where the half-life of the excited level ($E_1$) is 234~ps~\cite{NuclearDataTables}. Figure~\ref{fig:alpha_Pu239_240} shows the alpha spectrum for this sample, constructed in the approximation of only positional noise at $T_{SQL}=42$~nK, and with additionally accounted $\gamma-$decay. In this case, the SNR for each $\alpha-$peak is calculated individually with and without the associated gamma energy. The inset in Fig.~\ref{fig:alpha_Pu239_240} shows the SNR for the $\alpha-$peak, $E_\alpha=$5105.5~keV, calculated in the approximation of only positional noise at $T_{SQL}$, in the approximation of only $\gamma-$decay noise, and, when both effects are included. For particles smaller than $200$~nm, the total noise approaches the $\gamma-$decay noise asymptotically, entering the regime where it is independent of particle size or trap frequency, since $\alpha-$ and $\gamma-$kicks have a similar dependence on these parameters.

\begin{figure}[t]
	\includegraphics[width=0.49\textwidth]{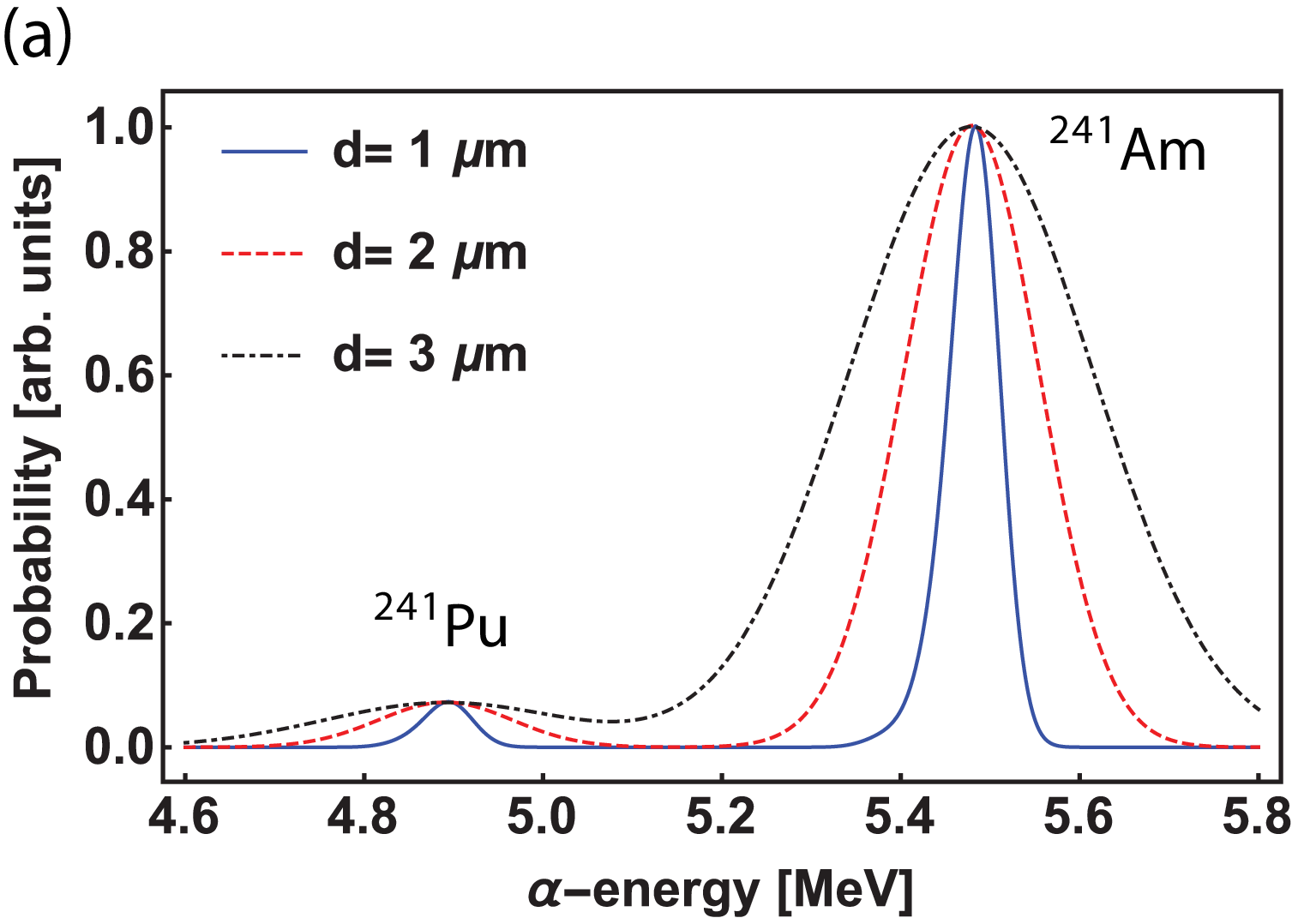}
	\includegraphics[width=0.49\textwidth]{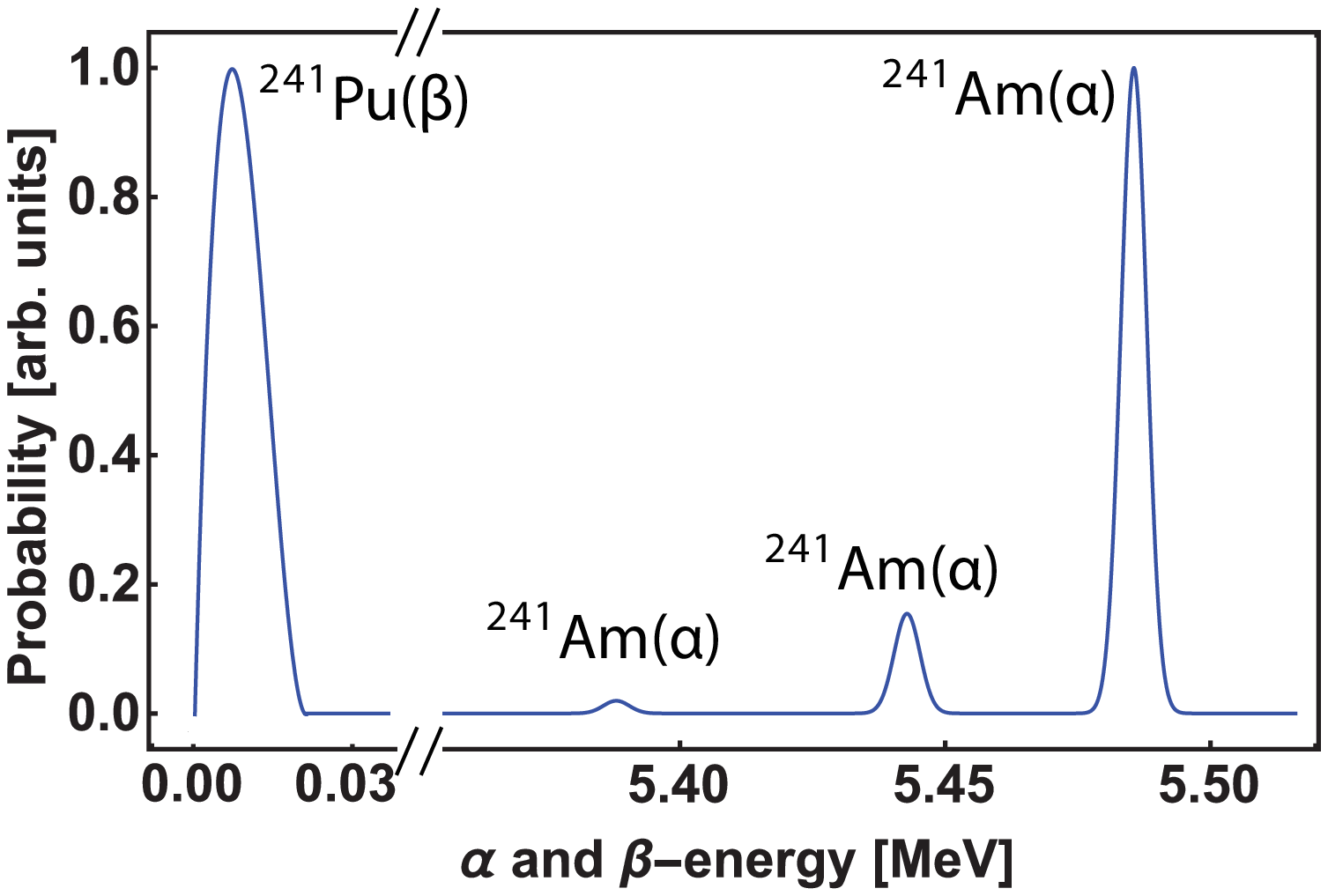}
	\caption{(a) $\alpha-$recoil spectrum for a recently purified $^{241}$Pu sample containing 1\% $^{241}$Am for various particle diameters. (b) $\alpha-$ and $\beta-$recoil spectrum for an aged $^{241}$Pu sample containing 4.5\% $^{241}$Pu and 86\% $^{241}$Am for $d=0.2$~$\mu$m. The beta part of the spectrum is plotted in the approximations of massless neutrino and no distortion by the Coulomb potential.} 
	\label{fig:alpha_beta}
\end{figure}

So far, we have discussed the applications of nuclear recoil spectroscopy for the study of samples composed of different isotopes of the same element. Another possible application of this method is obtaining the ratio of the elements in a sample belonging to the same nuclear decay chain, which would allow the determination of the age of the sample. For example, consider a hypothetical sample of recently purified $^{241}$Pu (half-life  $\sim$14.3 years). This isotope beta-decays to $^{241}$Am with a probability of 99.998\% as described in equation~(\ref{eq:beta_Pu241}), and alpha-decays with a probability of only 0.00247\%, according to,
\begin{equation}\label{Eq:alpha_Pu241}
^{241}\mbox{Pu}\rightarrow\;^{237}\mbox{U}^{2-}+\alpha^{2+},
\end{equation}
The most frequent decay product for $^{241}$Pu is $^{241}$Am, which in turn $\alpha-$decays to $^{237}$Np according to equation~(\ref{alpha_for_gamma}). Figure~\ref{fig:alpha_beta} (a) presents the $\alpha-$decay spectrum constructed from particle recoil for a sample of $^{241}$Pu with 1\% ingrown $^{241}$Am, which corresponds to an age of $\sim76$ days since last purification. The spectrum is plotted for particles of different sizes, and the widths are defined by the positional noise at $T_{SQL}=0.42~\mu$K in a 9 kHz optical trap. The effects of $\gamma-$broadening are not included in the calculations, since they are negligible in comparison to the positional noise broadening for these large particle sizes. It takes on average $\sim$33 min, $\sim$4~min, and $\sim$1~min for 1~$\mu$m, 2~$\mu$m, and 3~$\mu$m size particles, respectively, for one alpha decay of $^{241}$Pu in this sample to occur.  As shown in Fig.~\ref{fig:alpha_beta} (a), the peak widths depend strongly on the particle size. This suggests that selecting a smaller particle is advantageous for obtaining a better resolution, and hence, a better accuracy for extracting the ratio of the isotopes of the sample. On the other hand, in order to obtain proper statistics of the displacements for a short time interval in comparison to the accuracy of the age determination, it would be beneficial to analyze a bigger particle. Therefore, there is a trade-off between the analysis time and energy resolution, which can be mitigated by lowering trap frequencies and cooling the COM motion to lower temperatures.  

For comparison, consider an aged sample of $^{241}$Pu (4.5\%) with ingrown $^{241}$Am (86\%), which corresponds to an age of 64 years since last purification. For this sample, $\alpha$-decays in $^{241}$Pu are very rare in comparison to $^{241}$Am decays, and therefore, alpha-recoil spectrometry is not able to provide an accurate isotope composition within a reasonable time. However, our method is capable of measuring both alpha and beta recoils simultaneously, which is useful for this example, because the decay of $^{241}$Pu is dominated by beta decay. Figure~\ref{fig:alpha_beta} (b) depicts the combined alpha and beta decay spectrum for this sample for a 0.2~$\mu$m particle at the SQL temperature (0.42~$\mu$K) in a 9~kHz trap. Note that the beta peak is asymmetric. For simplicity, it is plotted in the approximation of massless neutrino, ignoring both the distortion by the Coulomb potential~\cite{BetaDecayCh8}, and the spectrum broadening due to positional noise. We speculate that the beta spectrum can be reconstructed from the proposed experiment as far as the positional noise ($\sim17$~pm at $T_{SQL}$ in a 9 kHz trap) is less than the recoil displacement ($\sim29$~pm) for a $\beta-$particle with mean $\beta-$energy (5.2~keV).  %may be mention other materials/examples. 

\section{Discussion}\label{discuss}
We expect that nuclear recoil spectroscopy of optically levitated particles, as proposed here, will be beneficial to the analysis of small particle samples, ranging from several nanometers to several micrometers, a type of analysis that is difficult by common mass spectrometry techniques. In principle, our approach can be universally applied to the analysis of any radioactive sample, as long as the activity and size of the particle lie within the limits described here. The proposed technique does not require the use of specific detectors for each type of emitted radiation, because it does not rely on the detection of emitted particles, as in conventional alpha/beta/gamma spectrometry.   

Having established that it is possible to detect and analyze individual displacements due to nuclear recoil in our system, we turn our attention to other possible applications aside from a nuclear recoil spectrometer. First, the possibility to resolve each individual recoil without actual detection of the emitted nuclear particle will allow the measurement of decay rates, and shed light into the ``uncertainty of the half-life" debate~\cite{Pomme1}. Oscillations in the decay rate are predicted by the theory of quantum mechanics~\cite{Khalfin,Nature_note_decay}. These oscillations are expected to happen only at very short time scales (Zeno and anti-Zeno effects~\cite{Zeno}), or at very long time scales ~\cite{Rothe}, with respect to the half-life. Some violations of these predictions have been observed at the very short time scale~\cite{Experiment_short_time_decay}. In addition, some research groups have reported on the observation of seasonal oscillations in the measured decay constant at time scales of the order of the half-life~\cite{Alburger,Siegert}, which have been potentially explained as the effects of neutrino impacts on the nuclear decay process, because the neutrino flux varies depending on the distance from the Sun to the Earth~\cite{Fischbach}, and also by the effects of cosmic neutrinos from dark matter~\cite{Parkhomov}. Controversially, other research groups have explained these oscillations as errors on the ionization chamber measurements used for detection of nuclear particles~\cite{Pomme2,Pomme3,Semkow}. They suggest that these errors experience seasonal oscillations due to temperature fluctuations within the laboratory environment. Such temperature fluctuations do not seem to be problematic for our method of  nuclear recoil detection. %\gamma attenuation in the ionization chamber becomes larger with increasing the dencity, which is proportionally increased with temperature growth [Semkow]...
     
In addition, it may soon be possible to study plutonium particle migration, which is observed routinely during plutonium operations, but has not been conclusively explained yet  \cite{PseudoEvaporation}. Specifically, long distance migration of plutonium metal and plutonium oxide particles cannot be fully explained by recoil alone. It is more likely to be caused by either particle fragmentation, or evaporation due to the high temperature of the bulk. The experimental scheme proposed here may be able to aid in providing an explanation of this phenomenon.

With regards to fundamental physics, we speculate that our approach may be of interest in the area of neutrino research. As discussed in section~\ref{TM}, the presence of neutrinos can clearly be observed during a beta decay process. Currently, there is a great interest in measuring the neutrino mass~\cite{NeutrinoMassReview1,NeutrinoMassReview2,RaizenHigh}, and in searching for neutrino-less double-$\beta$ decay~\cite{Majorana,GrattaNeutrino,MajoranaFresh}. However, despite the expected high resolution measurements of displacements due to recoil, our proposed system does not detect beta and neutrino emissions independently. An additional way of measuring the energy and momentum of a beta particle would be necessary for this purpose~\cite{commenttoRaiz}.

%4)   Trapping different types of particles: oxides, metals, etc or sample preparation...  
%
%Strictly speaking, the motion in such oscillator can be coupled along different axes, and strongly depends on the experimental realization of the optical trap (trap configuration; one or several beams; different regimes: particle size in comparison to the laser wavelength, etc.). However, there are regimes, when the motions along different axes are uncoupled from each other. For example, in such a regime the particle Brownian motion was successfully studied and its instantaneous velocities were measured\ \cite{T_Li_Brownian}.  Hence, for simplicity, we will further use one-dimensional model of an optical trap for understanding and quantifying our concept.
%
%The techniques to reach the standard quantum limit are actively studied at the moment due to interest of cooling macroscopic object to the quantum ground state and beyond that by applying fundamentally different trapping approaches \cite{IonTrap1,IonTrap2}  

\section{Conclusion}
In this work, we have demonstrated that it should be possible to detect and analyze nuclear decay events in an optically levitated particle. We have pointed out how the sensitivity of this detection system scales with the typical parameters of a realistic experiment. We have also discussed the possible application of this approach to a novel method of nuclear recoil spectroscopy, in particular, of those materials which are of interest to nuclear forensics. 

\section{Acknowledgments}
We thank the Chemistry Division of Los Alamos National Laboratory for providing program development funds, and Dr.~Joshua Bartlett for helpful discussions.

%%%%%%%%%%%%%%%%%%%%%%%%%%%%%%%%%%%%%%%%%%%%%%%%%%%%%%%%%%%%%%%%%%%%%%%%%%%%%%%%%%%%%%%%


\begin{thebibliography}{99}
\bibitem{Comment} The experimentally reachable temperatures along each axis depend on particle size, trap geometry, and also on the particular position diagnostics and cooling methods employed. Thus, the temperatures along different axes may vary. However, here we neglect these dependencies for simplicity.
\end{thebibliography}
\end{document}